\documentclass[aps,twocolumn,preprintnumbers,amsmath,amssymb,floatfix,superscriptaddress]{revtex4}
\usepackage{graphicx}
\usepackage{color}
\def\tHp{\tau_{\rm Hp}}

\def\SHp{S_{\rm Hp}}
\def\RHp{Re_{\rm Hp}}

\def\qMin{q_{\rm min}}
\def\qRes{q_{\rm s}}
\def\xs{x_{\rm s}}

\def\rsi{r_{{\rm s}i}}

\def\rsa{r_{\rm s1}}
\def\rsb{r_{\rm s2}}

\def\mPeak{m_{\rm peak}}
\def\gPeak{\gamma_{\rm peak}}
\def\gDrive{\gamma_{\rm drive}}

\def\flPsi{\widetilde{\psi}}
\def\flPhi{\widetilde{\phi}}

\def\eqPsi{\overline{\psi}}
\def\eqPhi{\overline{\phi}}
\def\eqVor{\overline{u}}
\def\eqCur{\overline{j}}

\def\gLin{\gamma_{\rm lin}}
\def\gKin{\gamma^{\rm kin}}
\def\gMag{\gamma^{\rm mag}}
\def\EKin{E^{\rm kin}}
\def\EMag{E^{\rm mag}}

\def\Ma{M^{(1)}}
\def\Mb{M^{(2)}}

\definecolor{gray}{rgb}{0.5,0.5,0.5}
\definecolor{dred}{rgb}{0.5,0.0,0.0}
\definecolor{dgreen}{rgb}{0.0,0.5,0.0}
\definecolor{dblue}{rgb}{0.0,0.0,0.5}

\begin{document}

\preprint{physics/06010XX}

\title{Dynamics of resistive double tearing modes with broad linear spectra}

\author{Andreas~Bierwage}
\email{abierwag@uci.edu}
\affiliation{Graduate School of Energy Science, Kyoto University, Gokasho, Uji, Kyoto 611-0011, Japan}
\altaffiliation[Present address: ]{Department of Physics and Astronomy, University of California, Irvine, CA 92697, USA}
\author{Sadruddin~Benkadda}
\email{benkadda@up.univ-mrs.fr}
\affiliation{Equipe Dynamique des Syst\`{e}mes Complexes, UMR 6633 CNRS-Universit\'{e} de Provence, 13397 Marseille, France}
\author{Satoshi~Hamaguchi}
\email{hamaguch@ppl.eng.osaka-u.ac.jp}
\affiliation{Center for Atomic and Molecular Technologies, Osaka University, 2-1 Yamadaoka, Suita, Osaka 565-0871, Japan}
\author{Masahiro~Wakatani}
\thanks{deceased}
\affiliation{Graduate School of Energy Science, Kyoto University, Gokasho, Uji, Kyoto 611-0011, Japan}

\date{\today}

\begin{abstract}
The nonlinear evolution of resistive double tearing modes (DTMs) with safety factor values $q=1$ and $q=3$ is studied with a reduced cylindrical model of a tokamak plasma. We focus on cases where the resonant surfaces are a small distance apart. Recent numerical studies have shown that in such configurations high-$m$ modes are strongly unstable and may peak around $m = \mPeak \sim 10$. In this paper, it is first demonstrated that this result agrees with existing linear theory for DTMs. Based on this theory, a semi-empirical formula for the dependence of $\mPeak$ on the system parameters is proposed. Second, with the use of nonlinear simulations, it is shown that the presence of fast growing high-$m$ modes leads to a rapid turbulent collapse in an annular region, where small magnetic island structures form. Furthermore, consideration is given to the evolution of low-$m$ modes, in particular the global $m=1$ internal kink, which can undergo nonlinear driving through coupling to fast growing linear high-$m$ DTMs. Factors influencing the details of the dynamics are discussed. These results may be relevant to the understanding of the magnetohydrodynamic (MHD) activity near the minimum of $q$ and may thus be of interest for studies on stability and confinement of advanced tokamaks.
\end{abstract}

\maketitle

\thispagestyle{empty}

\section{Introduction}

The advanced tokamak (AT) scenario, where the maximum current density is located off the magnetic axis, is of considerable interest for achieving thermonuclear fusion conditions and a quasi-steady-state operation in future tokamak devices (e.g., Refs.~\cite{Kikuchi93, Goldston94}). The associated non-monotonic $q$ profiles have pairs of resonant surfaces with the same rational value $\qRes = m/n$ and may give rise to double tearing modes (DTMs) \cite{Furth73}. Several detailed studies of DTMs were motivated by their possible role in rapid current penetration, compound sawtooth oscillations, off-axis sawtooth crashes and disruptions (e.g., Ref.~\cite{Bierwage05b} and references therein). It is important to understand the behavior of these instabilities in order to ensure efficient profile control and safe operation of tokamak devices.

In this paper we consider cases with $\qRes = 1$ and $\qRes > 1$ DTMs, both of which have attracted much attention in experiments. For instance, compound sawtooth oscillations were observed in Tokamak Experiment for Technology Oriented Research (TEXTOR) after two $\qRes = 1$ resonances had formed \cite{Koslowski97}. Off-axis sawteeth were observed on Tokamak Fusion Test Reactor (TFTR) when the minimum of the safety factor, $\qMin$, is near or below 2 \cite{Chang96}. Resistive and neoclassical $\qRes = 2$ DTMs were investigated in Axisymmetric Divertor Experiment (ASDEX)-Upgrade, one motivation being the possible interaction of these modes with internal transport barriers (ITBs) \cite{Guenter99, Guenter00}. In Japan Atomic Energy Agency Tokamak Upgrade (JT-60U) $\qRes = 3$ DTMs are thought to play a crucial role in disruptions, as indicated by experimental observations and numerical results \cite{Ishii00, Ishii02, Takeji02}.

In contrast to previous numerical studies, which focused mostly on cases where two resonant surfaces are located a relatively large distance apart (e.g., Ref.~\cite{Ishii03a}) and the linearly most unstable mode has the lowest possible poloidal mode number [e.g., $(m,n)=(3,1)$ for $\qRes = 3$], our interest lies in the regime where the distance between neighboring resonant surfaces is still small. For this case, it was found that DTMs (and multiple tearing modes in general) with high poloidal and toroidal mode numbers $m$ and $n$ are strongly unstable \cite{Bierwage05a}. The linear instability of DTMs for equilibria with small inter-resonance distances was studied numerically in Ref.~\cite{Bierwage05b}. The purpose of the present paper is (i) to shown that the findings of Ref.~\cite{Bierwage05b} agree with linear DTM theory, and (ii) to present first nonlinear simulation results involving high-$m$ DTMs for cases with $\qRes = 1$ and $\qRes > 1$. For simplicity, a reduced magnetohydrodynamic (RMHD) model is employed.

Existing linear theory for DTMs \cite{Pritchett80} predicts that in the strongly coupled limit (small mode numbers, small inter-resonance distance) the linear growth rate increases with the mode number as $\gLin \propto m^{2/3}$. Here, it is shown that this scaling agrees with numerical results. A semi-empirical analytical formula for estimating the mode number $\mPeak$ of the fastest growing mode in terms of the system parameters is proposed, based on the transition criterion between the strongly and weakly coupled limits derived in Ref.~\cite{Pritchett80}.

The magnetic island dynamics during the nonlinear evolution of $\qRes = 3$ and $\qRes = 1$ DTMs is described. In the $\qRes = 1$ case, the $m=1$ internal kink mode is unstable and eventually dominates the dynamics. However, full reconnection is not possible for small inter-resonance distances, so the core merely undergoes oscillatory motion. Full reconnection is only possible when the inter-resonance distance has already increased beyond the limit where modes with $m>1$ become sub-dominant or negligible compared to the $m=1$ mode. This is demonstrated using an intermediate case where the $m=2$ mode has a slightly higher growth rate than the $m=1$ mode.

Finally, the early evolution of the $m=1$ mode is investigated in detail. Of particular interest is the nonlinear driving due to fast growing high-$m$ DTMs. It is shown how the efficiency of this driving depends on the initial conditions of the simulation and that the driving, despite its radial localization, is capable of triggering the global resistive $m=1$ internal kink mode, similarly to the case of $\qRes = 1$ triple tearing modes (TTMs) studied recently \cite{Bierwage06a}.

This paper is organized as follows. In Section~\ref{sec:model} the physical model is introduced. In Section~\ref{sec:equlib-lin} we describe the equilibrium configurations used and their linear dispersion relation. Section~\ref{sec:analysis} is dedicated to a comparison between linear theory and numerical data, and in Section~\ref{sec:results} nonlinear simulation results are presented. In Section~\ref{sec:conclude} we draw conclusions, discuss possible applications and motivate further research in this direction.

\section{Model}
\label{sec:model}

We use the reduced magnetohydrodynamic (RMHD) equations in cylindrical geometry and in the limit of zero beta \cite{Strauss76, NishikawaWakatani}. This model has proven to be useful in studies of MHD instabilities, when the focus is on a qualitative description of fundamental aspects of the magnetized plasma system, as is the case here. The RMHD model governs the evolution of the magnetic flux function $\psi$ and the electrostatic potential $\phi$, as described previously in Ref.~\cite{Bierwage05b}. The normalized RMHD equations are
\begin{eqnarray}
\partial_t\psi &=& \left[\psi,\phi\right] - \partial_\zeta\phi - \SHp^{-1}\left(\hat{\eta}j - E_0\right)
\label{eq:rmhd1}
\\
\partial_t u &=& \left[u,\phi\right] + \left[j,\psi\right] + \partial_\zeta j + \RHp^{-1}\nabla_\perp^2 u.
\label{eq:rmhd2}
\end{eqnarray}

\noindent The time is measured in units of the poloidal Alfv\'{e}n time $\tHp = \sqrt{\mu_0 \rho_{\rm m}} a/B_0$ and the radial coordinate is normalized by the minor radius $a$ of the plasma. $\rho_{\rm m}$ is the mass density and $B_0$ the strong axial magnetic field. The current density $j$ and the vorticity $u$ are related to $\psi$ and $\phi$ through $j = -\nabla_\perp^2\psi$ and $u = \nabla_\perp^2\phi$, respectively.

Resistive diffusion is measured by the magnetic Reynolds number $\SHp = \tau_\eta / \tHp$ in Eq.~(\ref{eq:rmhd1}), with $\tau_\eta = a^2\mu_0/\eta_0$ being the resistive diffusion time and $\eta_0=\eta(r=0)$ the electrical resistivity at $r=0$. We use $\SHp = 10^6$, which is numerically efficient and physically reasonable in the framework of the model used.
Viscous dissipation is measured by the kinematic Reynolds number $\RHp = a^2/\nu\tHp$ in Eq.~(\ref{eq:rmhd2}), where $\nu$ is the kinematic ion viscosity. We choose regimes where the Prandtl number $Pr = \SHp/\RHp$ satisfies $Pr \sim 10^{-2}$, so that the viscosity effect is limited to small-scale flows and does not affect the instability of the dominant modes.

The source term $\SHp^{-1}E_0$ in Eq.~(\ref{eq:rmhd1}) compensates the resistive diffusion of the equilibrium current profile on the time scale $\tau_{\rm R} / \tHp = D_{12}^2\SHp$, where $D_{12}$ is the length scale of interest, i.e., here the inter-resonance distance (normalized by $a$). $E_0$ is taken to be constant, so the resistivity profile is given in terms of the equilibrium current density distribution as $\hat{\eta}(r) = \eqCur(r=0)/\eqCur(r)$. For simplicity, the temporal variation of the resistivity profile $\hat{\eta}$ is neglected.

As in Ref.~\cite{Bierwage05b}, each field variable $f$ is decomposed into an equilibrium part $\overline{f}$ and a perturbation $\widetilde{f}$ as
\begin{equation}
f(r,\vartheta,\zeta,t) = \overline{f}(r) + \widetilde{f}(r, \vartheta, \zeta, t).
\end{equation}

\noindent The system is described in terms of the Fourier modes, $\psi_{m,n}$ and $\phi_{m,n}$, obtained from the expansion
\begin{equation}
f(r, \vartheta, \zeta, t) = \frac{1}{2}\sum_{m,n} f_{m,n}(r,t).e^{i(m\vartheta - n\zeta)} + {\rm c.c.},
\end{equation}

\noindent with $m$ being the poloidal mode number and $n$ the toroidal mode number. In the following, the $(m,n)$ subscripts will often be omitted for convenience. We consider only the nonlinear couplings between modes of a single helicity $h=m/n$, so the problem is reduced to two dimensions. Results for the linearized system are obtained using initial-value and eigenvalue solvers as described in Ref.~\cite{Bierwage05b}. The nonlinear RMHD equations are solved numerically using the simulation code described in Ref.~\cite{Bierwage06a}.

In order to ensure numerical accuracy, we have evaluated the energy balance (temporal change in the system's energy compared to the dissipated energy) and compared the results obtained with different numbers of grid points and Fourier modes. In particular, the time histories and mode structures of individual modes were inspected in detail. Both linear and nonlinear calculations were performed with a grid spacing of $\Delta r = 5\times 10^{-4}$ and smaller. The number of Fourier modes is specified below for each case. In the regimes where the linear theory of DTMs \cite{Pritchett80} is valid, it was used to benchmark numerical results (e.g., the resistivity scaling $S_{\rm Hp}^\alpha$ with $\alpha \in [1/3,3/5]$; cf.~Figs.~11--14 in Ref.~\cite{Bierwage05b}).

\section{Equilibrium and linear instability}
\label{sec:equlib-lin}

The equilibrium state is taken to be axisymmetric (only $m=n=0$ components) and free of flows, i.e., $\eqPhi = \eqVor = 0$. The equilibrium magnetic configuration is uniquely defined in terms of the safety factor $q(r)$, and the magnetic flux function and current density profiles are obtained though the relations
\begin{equation}
q^{-1} = -\frac{1}{r} \frac{{\rm d}}{{\rm d}r}\psi_{0,0} \quad {\rm and} \quad j_{0,0} = \frac{1}{r} \frac{{\rm d}}{{\rm d}r} \frac{r^2}{q}.
\label{eq:q-equlib}
\end{equation}

\noindent The model equation used for the $q$ profile is \cite{Bierwage05b}
\begin{equation}
q(r) = q_0.F_1(r).\left\{1 + \left(r/r_0\right)^{2 w(r)}\right\}^{1/ w(r)},
\label{eq:qModel}
\end{equation}

\noindent where
\begin{eqnarray}
r_0 &=& r_{\rm A}\left|\left[m/(nq_0)\right]^{w(r_{\rm A})} - 1\right|^{-1/[2 w(r_{\rm A})]}, \nonumber
\\
w(r) &=& w_0 + w_1r^2, \nonumber
\\
F_1(r) &=& 1 + f_1\exp\left\{-\left[(r - r_{11})/r_{12}\right]^2\right\}. \nonumber
\end{eqnarray}

\begin{figure}
[tb]
\centering
\includegraphics[height=6.0cm,width=8.0cm]
{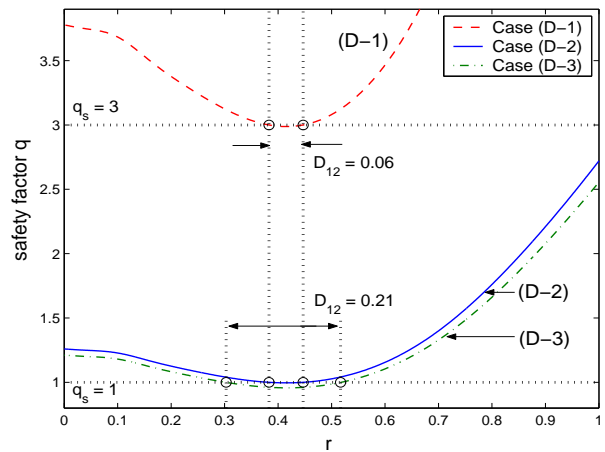}
\caption{(Color online). Equilibrium safety factor profiles $q(r)$ for Cases (D-1)--(D-3). The properties of these profiles are listed in Table~\protect\ref{tab:equlib} and dispersion relations are shown in Fig.~\protect\ref{fig:spec}.}
\label{fig:equlib_q}%
\end{figure}

\begin{figure}
[tb]
\centering
\includegraphics[height=6.0cm,width=8.0cm]
{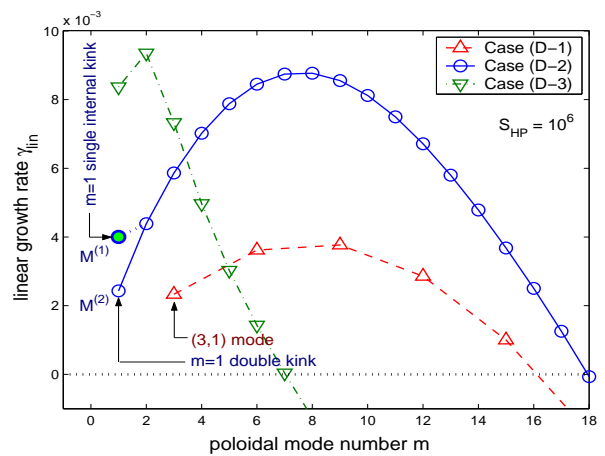}
\caption{(Color online). Spectra $\gLin(m)$ of unstable DTMs for the three cases in Fig.~\protect\ref{fig:equlib_q} for $\SHp = 10^6$ and $\RHp = 10^8$ [$\RHp = 10^7$ in Case (D-3)]. For Case (D-2) (circles) the growth rates of both $m=1$ eigenmodes, $\Ma$ (single kink) and $\Mb$ (double kink), are shown: $\gamma^{(1)}(m=1) = 4.0\times 10^{-3}$ and $\gamma^{(2)}(m=1) = 2.5\times 10^{-3}$. All other growth rates belong to $\Mb$-type eigenmodes, as defined in the text.}
\label{fig:spec}%
\end{figure}

\noindent With the parameter values in Table~\ref{tab:equlib_q-parm} the equilibrium $q$ profiles shown Fig.~\ref{fig:equlib_q} are produced, each of which has two resonant surfaces with $\qRes \equiv q(\rsi) = m/n$ at the radii $r=\rsi$ ($i=1,2$). The distance between the resonances, $D_{12} = |\rsb - \rsa|$ was chosen sufficiently small, so that broad spectra of DTMs are unstable, with dominant modes having $m$ greater than the lowest poloidal mode number that is consistent with a given field line pitch $\qRes = m/n$. The dispersion relations (spectra of linear growth rates) $\gLin(m)$ are plotted in Fig.~\ref{fig:spec}. In Cases (D-1) and (D-2), pairs of resonant surfaces with $\qRes = 3$ and $1$, respectively, are located a small distance $D_{12} = 0.06$ apart, so that the fastest growing mode has $\mPeak = 8$. In addition, we consider Case (D-3) with $qRes=1$, which has a larger inter-resonance distance, $D_{12} = 0.21$. Here, the dominant mode is $\mPeak = 2$. The characteristics of all cases are summarized in Table~\ref{tab:equlib}. Linear eigenmode structures of DTMs with various mode numbers $m$ were presented in Ref.~\cite{Bierwage05b}. Note that there can be up to two unstable eigenmodes for a given $(m,n)$. Eigenmodes peaking in the region $0 < r < \rsa$ (for $\qRes = 1$) or near $\rsa$ (for $\qRes > 1$) are denoted by $\Ma$, while those extending to the outer resonant surface are labeled $\Mb$. When $D_{12}$ is small, the $\Mb$-type eigenmodes are usually dominant for $m>2$ and only their growth rates are plotted in Fig.~\ref{fig:spec}.

\begin{figure}
[tb]
\centering
\includegraphics[height=6.0cm,width=8.0cm]
{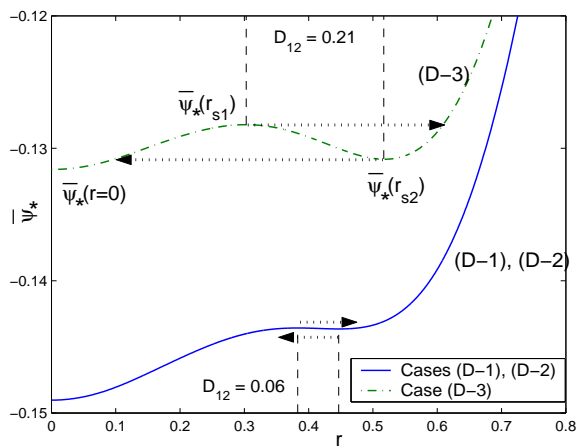}
\caption{(Color online). Equilibrium helical flux functions $\eqPsi_*(r)$ for Cases (D-2) and (D-3). The radial extent of ``reconnectable'' regions is indicated by arrows.}
\label{fig:equlib_psi}%
\end{figure}

Finally, a comment is in place with regard to the region that can undergo magnetic reconnection in the cases studied here. The profiles of the equilibrium helical flux functions $\eqPsi_* = \eqPsi + r^2/(2\qRes)$ in Fig.~\ref{fig:equlib_psi} show that the reconnectable regions do not include the magnetic axis at $r=0$. Since all cases satisfy $\eqPsi_*(r=0) < \eqPsi_*(\rsb)$, the reconnection can only be partial \cite{Parail80, Pfeiffer85}.

\begin{table*}
[tb]
\centering
\caption{Parameter values for the $q$ profiles in Fig.~\protect\ref{fig:equlib_q} using the model formula (\protect\ref{eq:qModel}).}
\label{tab:equlib_q-parm}
\begin{ruledtabular}
\begin{tabular}[t]{cccccccccc}
Case & $q_0$ & $r_{\rm A}$ & $w_0$ & $w_1$
& $m$ & $n$ & $f_1$ & $r_{11}$ & $r_{12}$ \\
\hline (D-1) & $2.6$ & $0.655$ & $3.8824$ & $0$ &
$3$ & $1$ & $-0.238$ & $0.4286$ & $0.304$ \\
(D-2) & $1.3$ & $0.655$ & $3.8824$ & $0$ &
$1$ & $1$ & $-0.238$ & $0.4286$ & $0.304$ \\
(D-3) & $1.25$ & $0.655$ & $3.8824$ & $0$ &
$1$ & $1$ & $-0.238$ & $0.4286$ & $0.304$ \\
\end{tabular}
\end{ruledtabular}
\end{table*}

\begin{table*}
[tb]
\centering
\caption{Properties of the $q$ profiles shown in Fig.~\protect\ref{fig:equlib_q}. The linear instability characteristics of these cases were previously studied in Ref~\protect\cite{Bierwage05b}, with the case labels given in the second column. The values of the magnetic shear and the resistivity at the resonant surfaces are denoted by $s_i \equiv s(\rsi)$ and $\hat{\eta}_i \equiv \hat{\eta}(\rsi)$, respectively. The mode numbers of the fastest growing mode, $\mPeak$, are valid for $\SHp = 10^6$ and $\RHp = 10^8$ [$\RHp = 10^7$ in Case (D-3)].}
\label{tab:equlib}
\begin{ruledtabular}
\begin{tabular}{cccccccccc}
Case & Case in Ref.~\protect\cite{Bierwage05b} & $\qRes$ &
$q_{\rm min}$ & $D_{12}$ & $s_1$ & $s_2$ &
$\hat{\eta}_1$ & $\hat{\eta}_2$ & $\mPeak$
\\
\hline (D-1) & (IIIb) & $3$ & $1.99$ & $0.06$ & $-0.10$ & $0.12$ &
$0.76$ & $0.84$ & $8$ \\
(D-2) & (Ia) & $1$ & $0.99$ & $0.06$ & $-0.10$ & $0.12$ &
$0.76$ & $0.84$ & $8$ \\
(D-3) & (Ib) & $1$ & $0.96$ & $0.21$ & $-0.20$ & $0.45$ &
$0.75$ & $1.07$ & $2$ \\
\end{tabular}
\end{ruledtabular}
\end{table*}

\section{Comparison with linear theory}
\label{sec:analysis}

Pritchett, Lee and Drake (PLD) \cite{Pritchett80} developed a linear theory for resistive DTMs by applying the techniques and results from the theory for the resistive $m=1$ internal kink mode by Coppi \textit{et~al.}~\cite{Coppi76}. PLD derived analytical forms for the linear growth rate $\gLin$ of DTMs in the limit of strong and weak coupling. The strongly coupled limit applies to modes which are close to ideal-MHD marginal stability, whereas weakly coupled modes are strongly ideal-MHD stable. Defining the poloidal wave number $k_\vartheta=m/r_0$ [$r_0 = (\rsa+\rsb)/2$] and the distance $\xs=D_{12}/2$, the validity of the dispersion relations in the two limits is given by the constraint \cite{Pritchett80}
\begin{equation}
\left(\frac{k_\vartheta^2}{B_{\rm s}' \SHp}\right)^{1/3} \ll \underbrace{k_\vartheta x_{\rm s}}\limits_{\rm strong} \ll \left(\frac{k_\vartheta^2}{B_{\rm s}' \SHp}\right)^{1/9} \ll \underbrace{k_\vartheta x_{\rm s}}\limits_{\rm weak} \ll 1
\label{eq:theory_valid}
\end{equation}

\noindent where $B_{\rm s}' = s(\rsi)/\qRes$, with $s=rq'/q$ being the magnetic shear evaluated at a resonant surface. The lower limit is based on the requirement that the resistive layer width $\delta_\eta$ be smaller than the distance $x_{\rm s}$, and the upper limit corresponds to a large-aspect-ratio approximation used in the derivation of the dispersion relation. Modes that fall into the transitional regime between the strongly and weakly coupled limits are neither close to marginal stability nor strongly ideal-MHD stable. In this section we reproduce essential steps from PLD's derivation and compare the theoretical predictions with numerical results, focusing in particular on the $m$ dependence of the linear growth rate. It is then shown that a semi-empirical formula for the mode number $\mPeak$ of the fastest growing mode can be extracted from the theory.

\begin{figure}
[tb]
\centering
\includegraphics[height=6.0cm,width=8.0cm]
{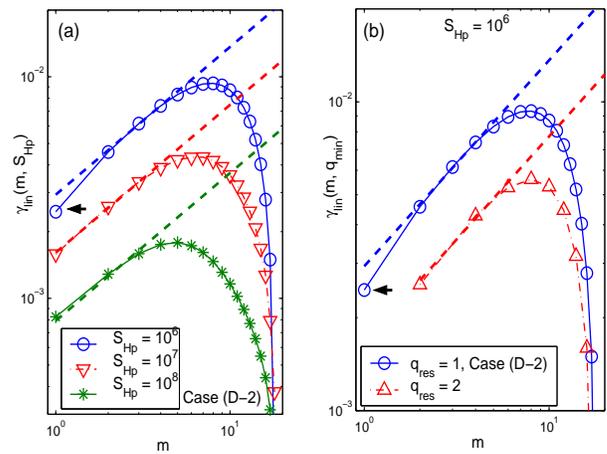}
\caption{(Color online). Comparison between theory and simulation: $m$ dependence of the linear DTM growth rate. Only the growth rates of $\Mb$-type modes are shown. The dashed lines indicate the scaling law $\gLin \propto m^{2/3}$. To fit Eq.~(\protect\ref{eq:power_law_strong}) to the data, $\hat{\lambda}$ in the range $0.7 \lesssim \hat{\lambda}_{\rm h} \lesssim 0.9$ is used. \\
(a): Spectra $\gLin^{(2)}(m, \SHp)$ of Case (D-2) for $\SHp = 10^6$, $10^7$, $10^8$. The data are the same as in Fig.~10 of Ref.~\protect\cite{Bierwage05b}. \\
(b): Spectra $\gLin^{(2)}(m, \qRes)$ for $\qRes = 1$ and $3$ [Cases (D-2) and (D-1)] with $\SHp = 10^6$. The data are the same as in Fig.~6(a) of Ref.~\protect\cite{Bierwage05b}.}
\label{fig:spec_comp}%
\end{figure}

The dispersion relation for strongly coupled DTMs (small $k_\vartheta x_{\rm s}$) is \cite{Pritchett80}
\begin{equation}
\gLin = \hat{\lambda} k_\vartheta^{2/3} B_{\rm s}^{\prime 2/3}\SHp^{-1/3} \propto m^{2/3} \quad \text{(strong coupling)},
\label{eq:power_law_strong}
\end{equation}

\noindent where $\hat{\lambda} = \hat{\lambda}(\hat{\lambda}_{\rm h})$ is a normalized growth rate. The dependence of $\hat{\lambda}$ on the ideal-MHD instability drive measured by $\hat{\lambda}_{\rm h}$ was obtained by PLD through asymptotic matching. At marginal stability one has $\hat{\lambda}_{\rm h} = 0$ and $\hat{\lambda}(0) = 1$. Equation~(\ref{eq:power_law_strong}) implies that the growth rate increases with the poloidal mode number $m$. A comparison of the $m^{2/3}$ power law with numerically computed growth rate spectra from Ref.~\cite{Bierwage05b} is given in Fig.~\ref{fig:spec_comp}. Good agreement is found, apart from the data point $\gLin(m=1)$ in Case (D-2) with $\SHp = 10^6$ [arrows in Figs.~\ref{fig:spec_comp}~(a) and (b)]. This deviation is most likely related to the fact that $\delta_\eta = [\gLin/(k_\vartheta^2 B_{\rm s}^{'2}\SHp)]^{1/4} \approx 0.37\times D_{12} = 0.74\times\xs$ in that case, which violates the condition $\delta_\eta \ll \xs$ \cite{Pritchett80}. Here, $\delta_\eta$ is the linear resistive layer width given by Eq.~(22) in Ref.~\protect\cite{Bierwage05b}. Note that in order to fit Eq.~(\ref{eq:power_law_strong}) to the numerical data in Fig.~\ref{fig:spec_comp} we use $\hat{\lambda} \lesssim 1$, which means that DTMs are only approximately marginally stabile and lie slightly in the ideal-MHD stable domain.

In the limit of weak coupling (large $k_\vartheta x_{\rm s}$) the dispersion relation is \cite{Pritchett80}
\begin{equation}
\gLin \simeq \left( \frac{8\Gamma(5/4)}{\gamma_{\rm h}\Gamma(-1/4)} \right)^{4/5} \left( \frac{k_\vartheta^2 B_{\rm s}^{\prime 2}}{\SHp} \right)^{3/5}.
\label{eq:disp_weak}
\end{equation}

\noindent The quantity $\gamma_{\rm h}$ is given by
\begin{equation}
\gamma_{\rm h} = -\frac{\pi k_\vartheta^3}{B_{\rm s}'} \int\limits_0^{\xs}{\rm d}x.B_*^2(x),
\label{eq:disp_ideal}
\end{equation}

\noindent and thus depends on the shape of the $q$ profile [here, $B_* = \overline{B}_\vartheta(r) - \overline{B}_\vartheta(\rsi)$ is the helical field that reverses sign across $r=\rsi$]. In the limit of $\xs \rightarrow 0$ one can approximate $B_*$ by a parabola centered half-way between the resonances, $B_* \approx B_{\rm s}'(\xs^2 - x^2)/2\xs$, which yields $\gamma_{\rm h} \approx B_{\rm s}'k_\vartheta^3 \xs^3$. With this, Eq.~(\ref{eq:disp_weak}) gives
\begin{equation}
\gLin \propto m^{-6/5} \quad \text{(weak coupling)},
\label{eq:power_law_weak}
\end{equation}

\noindent [Note that the parabolic approximation affects only the $\xs$ dependence, not the $m$ dependence. Here, we assume that $\xs$ be small and the weak coupling is realized through large $m$.] This result shows that in the weakly coupled limit the growth rate of DTMs decreases with increasing $m$. The $m^{-6/5}$ power law does not fit the data, which can already be seen from the convexity (${\rm d}^2\gLin/{\rm d}m^2 < 0$) of the spectra for the cases (D-1) and (D-2) in Fig.~\ref{fig:spec} as opposed to the concavity of $\gLin$ in Eq.~(\ref{eq:power_law_weak}). Here, it is useful to check the validity of the weakly coupled limit. Substitution of the profile parameters $x_{\rm s} \approx 0.03$, $r_0 = 0.42$ and $B_{\rm s}' \approx 0.11$ into Eq.~(\ref{eq:theory_valid}) we find $7 \ll m_{\rm weak} \ll 14$ ($\Delta m < 7$) for $\SHp = 10^6$. The fact that the $m^{-6/5}$ law is not observed in our data suggests that the transition regimes associated with the lower and upper bounds of $m_{\rm weak}$ are so broad that they overlap for $\Delta m < 7$ (note that the theory does not define the exact meaning of the symbol ``$\ll$''). For larger values of $\SHp$ the lower limit for $m$ is reduced. Indeed, the high-$m$ side of the spectrum for the cases (D-1) and (D-2) becomes concave for $\SHp = 10^8$ (cf.~Fig.~10 in Ref.~\cite{Bierwage05b}) where $5 \ll m_{\rm weak} \ll 14$ is required according to Eq.~(\ref{eq:theory_valid}). For case (D-3) where $x_{\rm s} \approx 0.1$, the weak coupling limit is valid for $1 \ll m_{\rm weak} \ll 4$ ($\SHp = 10^6$) which encompasses essentially all unstable modes as can be seen in Fig.~\ref{fig:spec}. Again, the number of unstable modes is too small for the spectrum to follow the $m^{-6/5}$ law in any significant interval. We conjecture that an appreciable number of modes satisfying the dispersion relation Eq.~(\ref{eq:disp_weak}) for the weakly coupled limit may only be found for very small $x_{\rm s}$ and very large $\SHp$.

One more comment is in place with regard to the upper boundary of the weak coupling limit, $k_\vartheta \xs \ll 1$. When this boundary is exceeded by increasing $\xs$ then weak coupling implies that the modes are spatially decoupled and eventually turn into independent single tearing modes. On the other hand, if $\xs$ is left sufficiently small, such that the perturbation at one resonant surface can still be ``felt'' at the other resonance, an increase in $k_\vartheta = m/r_0$ such that $r_0/m < \xs$ may be expected to weaken the mutual driving of the perturbations: an O-point at one resonance will ``push'' not only the facing X-point but also on O-points. To the best of the authors' knowledge, this regime is not yet understood. The stabilization of high-$m$ modes seems to depend on other factors, including the mechanism breaking the ideal-MHD constraint. For instance, collisionless DTMs due to electron inertia tend to have a broader spectrum of unstable modes than resistive DTMs for the same $q$ profile \cite{Bierwage06d}.

The growth rates plotted in Fig.~\ref{fig:spec} increase with $m$ for $m < \mPeak$ and a decrease for $m > \mPeak$. According to Eqs.~(\ref{eq:power_law_strong}) and (\ref{eq:power_law_weak}), this behavior corresponds to that predicted for the strongly and weakly coupled limit, respectively. Despite the lack of a distinguished band of weakly coupled modes in the cases considered here, we conjecture that an estimate for the mode number of the fastest growing mode $\mPeak$ may be obtained from the transition criterion in Eq.~(\ref{eq:theory_valid}) derived by PLD, i.e., $k_\vartheta x_{\rm s} \approx [k_\vartheta^2 / (B_{\rm s}' \SHp)]^{1/9}$. Solving this relation for $m$ yields
\begin{equation}
m_{\rm trans} \approx r_0 / (\xs^9 B_{\rm s}' \SHp)^{1/7}.
\label{eq:m-trans}
\end{equation}

\noindent for mode numbers in the transitional regime. Let us compare some values obtained from Eq.~(\ref{eq:m-trans}) with numerical results. For instance, for Case (D-2) (where $r_0 = 0.42$, $B_{\rm s}' \approx 0.11$, $\xs \approx 0.03$) one obtains $m_{\rm trans} = 7,5,4$ for $\SHp = 10^6,10^7,10^8$. The measured values for $\mPeak$ are $8,6,5$ (Fig.~\ref{fig:spec_comp}; see also Fig.~10 in Ref.~\protect\cite{Bierwage05b}), which suggests that $\mPeak \approx m_{\rm trans} + 1$. Tests with other configurations gave similarly good agreement, despite the fact that the $\xs$ dependence is described only approximately under the assumption that $q(r)$ is parabolic around $r=r_0$. Indeed, there is not much freedom for varying the shape of the $q$ profile in the inter-resonance region when the distance between the resonances must be small, so for realistic $q$ profiles the parabolic approximation may be expected to be sufficiently accurate.

Based on the good agreement between linear theory and simulation, we propose the semi-empirical formula
\begin{equation}
\mPeak \approx \frac{r_0}{(\xs^9 B_{\rm s}' \SHp)^{1/7}} + 1.
\label{eq:mpeak}
\end{equation}

\noindent for the dependence of $\mPeak$ on the system parameters. Equation~(\ref{eq:mpeak}) is useful for small $D_{12} = 2\xs$, where $\mPeak > 1$. Note that a small inter-resonance distance also implies that the difference between the magnetic shears $s_1$ and $s_2$ is small. Hence, due to the weak shear dependence it is not so important whether $\mPeak$ is evaluated using $s_1$ or $s_2$.

The poloidal mode number of the fastest growing linear mode, $\mPeak$, is useful for the interpretation of the nonlinear dynamics, since it determines the size of the magnetic island structures. Numerically computed values for the profiles used in this study are given in Table~\ref{tab:equlib}.

\section{Nonlinear results}
\label{sec:results}

\begin{figure}
[tbp]
\centering
\includegraphics[height=18.6cm]
{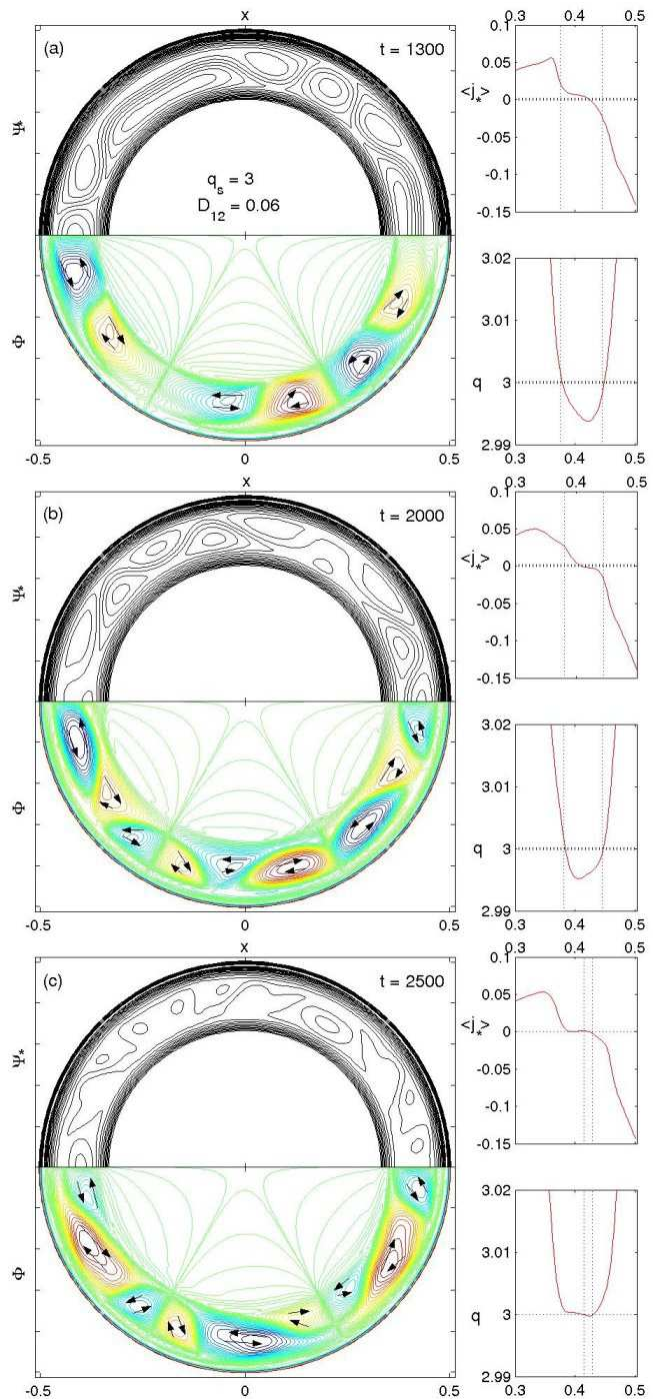} 
\caption{(Color online). Reconnection dynamics with $\qRes = 3$ DTMs for small inter-resonance distance $D_{12} = 0.06$ [Case (D-1)]. The snapshots were taken at (A) $t=1300$, (B) $t=2000$ and (C) $t=2500$. Each snapshot consists of contour plots of the helical flux $\psi_*$ (top) and the electrostatic potential $\phi$ (bottom). Arrows indicate the flow directions. On the right-hand side, the instantaneous profiles $q(r,t)$ and $\left<j_*\right> \equiv [j_*(r,t)]_{0,0}$ are shown. $\SHp = 10^6$, $\RHp = 10^8$.}
\label{fig:snaps_2tm_D-1}%
\end{figure}

\begin{figure}
[tbp]
\centering
\includegraphics[height=18.6cm]
{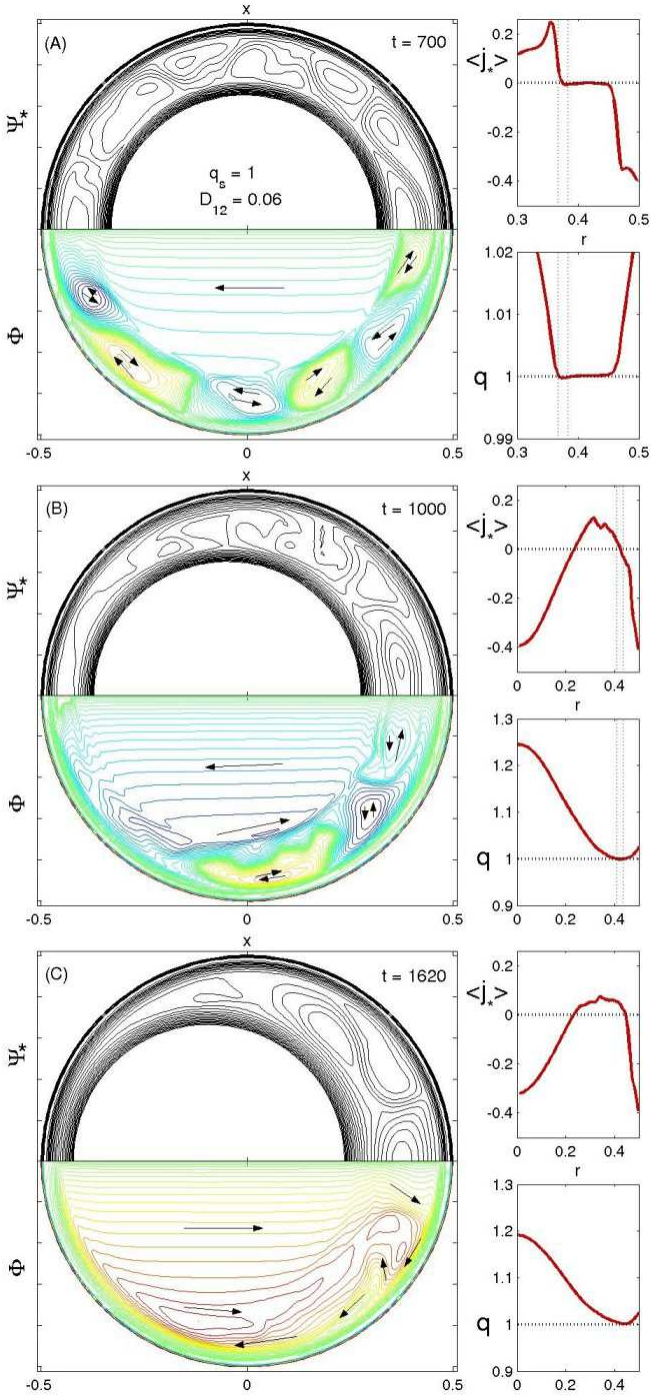} 
\caption{(Color online). Reconnection dynamics with $\qRes = 1$ DTMs for small inter-resonance distance $D_{12} = 0.06$ [Case (D-2)]. Arranged as Fig.~\protect\ref{fig:snaps_2tm_D-1}. $\SHp = 10^6$, $\RHp = 10^8$.}
\label{fig:snaps_2tm_D-2}%
\end{figure}

\begin{figure}
[tbp]
\centering
\includegraphics[height=18.6cm]
{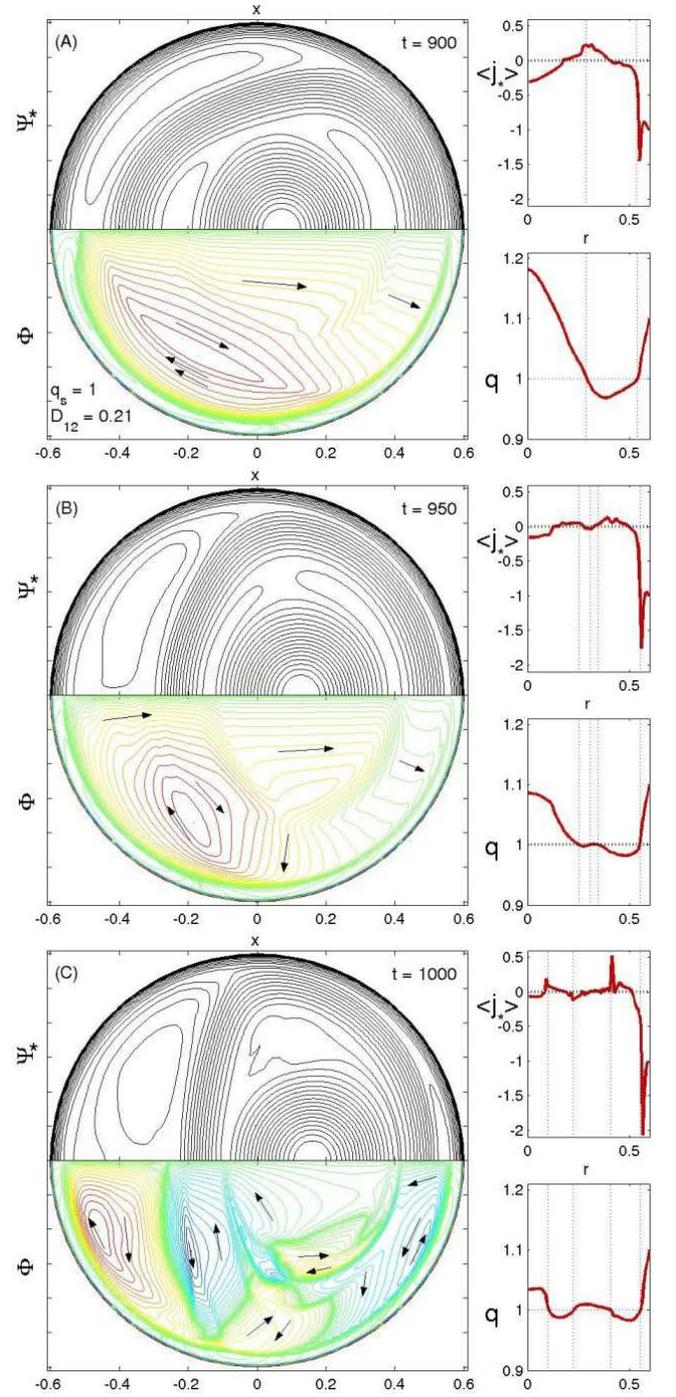} 
\caption{(Color online). Reconnection dynamics for $\qRes = 1$ DTMs with larger inter-resonance distance $D_{12} = 0.21$ [Case (D-3)]. Arranged as Fig.~\protect\ref{fig:snaps_2tm_D-1}. $\SHp = 10^6$, $\RHp = 10^7$.}
\label{fig:snaps_2tm_D-3}%
\end{figure}

Starting from an unstable equilibrium, the instability growth is excited by applying an initial perturbation of the form
\begin{equation}
\flPsi(t=0) = \frac{1}{2}\sum\limits_{m}\Psi_{0m} r (r-1) e^{i(m\vartheta_* + \vartheta_{0m})} + {\rm c.c},
\label{eq:pert}
\end{equation}

\noindent where $\Psi_{0m}$ is the perturbation amplitude (typically $10^{-7}$), $\vartheta_* \equiv \vartheta - \qRes^{-1}\zeta$ is a helical angle coordinate and $\vartheta_{0m}$ is an initial phase shift. The values $\vartheta_{0m} = 0$ and $\pi$ are assigned to each $m$ in a random manner. This introduces some degree of incoherence while retaining mirror symmetry about the $x$ axis (due to parity conservation in RMHD). This restriction improves numerical accuracy, simplifies visualization and has no significant effect on the central claims of this paper.

We begin with a description of the magnetic reconnection dynamics in Sec.~\ref{sec:results_islands} where the system as a whole is considered. In Sec.~\ref{sec:results_modes} the evolution of individual modes is analyzed in detail.

\subsection{Magnetic reconnection dynamics}
\label{sec:results_islands}

First, consider Case (D-1) where two $\qRes = 3$ resonances are located a small distance $D_{12} = 0.06$ apart. A nonlinear simulation was carried out using $\Psi_{0,m>0} = 10^{-7}$, $\SHp = 10^6$ and $\RHp = 10^8$, and including 32 modes ($m=0,3,...,93$). Snapshots showing contour plots of the helical flux function $\psi_* = \psi + r^2/(2\qRes)$ and the electrostatic potential $\phi$, as well as instantaneous profiles of $q(r)$ and $\left<j_*\right> \equiv [-\nabla_\perp^2\psi_*]_{0,0} = j_{0,0} - 2/\qRes$, are presented in Fig.~\ref{fig:snaps_2tm_D-1}. We observe that, in response to the random broad-band perturbation applied, magnetic reconnection occurs simultaneously at many locations, giving rise to a multitude of small magnetic islands. In Figs.~\ref{fig:snaps_2tm_D-1}(A) and (B) it can be seen that the dominant island sizes correspond to the mode numbers $m=6$ and $9$ (dispersion relation: $\mPeak = 9$). The onset of the reconnection is determined by the growth rate of the $(9,3)$ mode, which is about $1.6$ times that of the $(3,1)$ mode (cf.~Fig.~\ref{fig:spec}). The reconnection leaves behind an annularly flattened $q$ profile, as can be seen in Fig.~\ref{fig:snaps_2tm_D-1}(C).

Next, let us investigate the response of $\qRes = 1$ DTMs in Case (D-2). The calculation was performed with $\Psi_{0,m>0} = 10^{-7}$, $\SHp = 10^6$ and $\RHp = 10^8$, and 128 modes ($m=0,1,...,127$) were included. As can be seen in Fig.~\ref{fig:snaps_2tm_D-2}(A), the reconnection dynamics begin with an annular collapse with more or less turbulent patterns dominated by mode numbers around $m \sim 7$--$9$, in accordance with the peak of the linear dispersion relation ($\mPeak = 8$). Note in Fig.~\ref{fig:snaps_2tm_D-2}(A) that the $q$ profile has been flattened in the inter-resonance region and that the resistive $m=1$ internal kink mode is not yet involved in the dynamics. The kink appears at a later time, leading to a growing core displacement inside the turbulent region, as can be seen in Fig.~\ref{fig:snaps_2tm_D-2}(B). After all reconnectable flux surfaces have been reconnected a rebound occurs and the core displacement decays as indicated by the arrows in Fig.~\ref{fig:snaps_2tm_D-2}(C). The further evolution was not investigated, but continuing oscillation of the core is to be expected.

Finally, in Fig.~\ref{fig:snaps_2tm_D-3} simulation results are presented for Case (D-3). The calculation was performed with $\Psi_{0,m>0} = 10^{-7}$, $\SHp = 10^6$ and $\RHp = 10^7$, including 128 modes ($m=0,1,...,127$). Case (D-3) is a realization of the intermediate regime where two $\qRes = 1$ resonances are located so far apart that the fastest growing mode is $m=2$, closely followed by $m=1$. It can be seen that magnetic islands with dominant mode numbers $m=2$ and $m=1$ largely determine the structure of the magnetic surfaces [Fig.~\ref{fig:snaps_2tm_D-3}(A) and (B)]. In the present case, the $m=1$ mode tends to dominate on the outer resonance and $m=2$ on the inner one, which leads to a strong deformation of the inter-resonance region into a D-shape [Fig.~\ref{fig:snaps_2tm_D-3}(C)]. This calculation had to be terminated soon after snapshot~(C) due to a continuing increase in the energies of high-$m$ modes. An island separatrix is approaching the coordinate origin, $r=0$, and our numerical code is not suitable for further following dynamics with such a kind of asymmetry.

Equilibria with larger inter-resonance distance $D_{12}$ than in Case (D-3) have a dominant $m=1$ mode and the dynamics proceed as described in earlier studies (e.g., Refs.~\cite{Parail80, Ishii00}).

\subsection{Detailed evolution of individual modes}
\label{sec:results_modes}

\begin{figure}
[tb]
\centering
\includegraphics[height=7.960cm,width=8.0cm]
{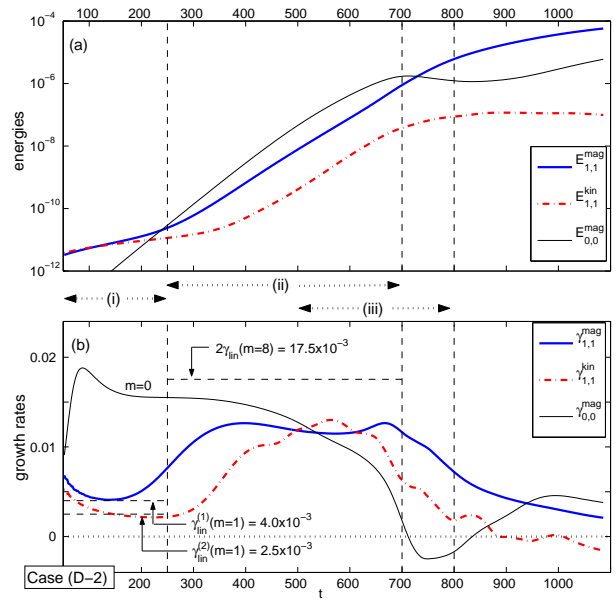}
\caption{(Color online). Evolution of the $m=1$ and $m=0$ modes in Case (D-2). In (a) the evolution of the magnetic and kinetic energies $\EMag$ and $\EKin$ [Eq.~(\protect\ref{eq:def-E})] is shown. In (b) the magnetic and kinetic growth rates $\gMag$ and $\gKin$ [Eq.~(\protect\ref{eq:def-gr})] are plotted. The main stages relevant to the evolution of the $m=1$ mode are: (i) establishing the linear mode structure and linear growth, (ii) nonlinearly driven growth, (iii) reconnection in the inter-resonance region (annular collapse). In diagram (b), the dashed horizontal line in phase (i) indicates the linear growth rate $\gLin(m=1) = 4.0\times 10^{-3}$ from Fig.~\protect\ref{fig:spec}. The dashed horizontal line during stage (ii) indicates the expected growth rate due to nonlinear driving. $\SHp = 10^6$, $\RHp = 10^8$, $\Psi_{0,m>0}=10^{-7}$, and the initial phases for modes $m=6$--$10$ are $\vartheta_{0m} = \{ 0,0,\pi,\pi,0 \}$. Similar behavior is also found for cases with $\qRes > 1$, such as (D-1).}
\label{fig:E-g_D-2}%
\end{figure}

\begin{figure}
[tb]
\centering
\includegraphics[height=7.273cm,width=8.0cm]
{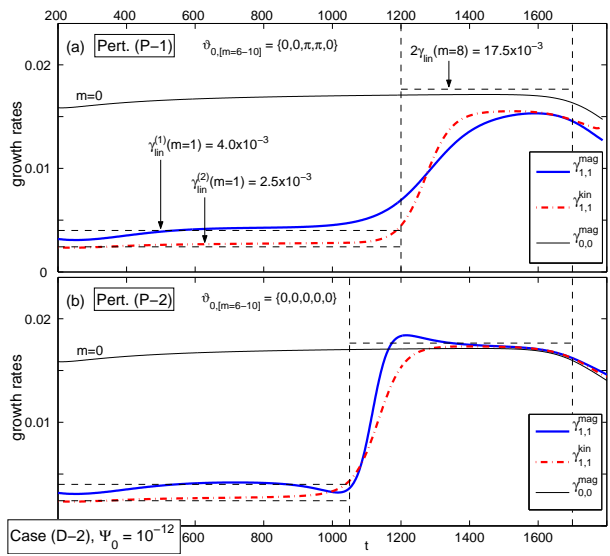}
\caption{(Color online). Evolution of the growth rates of the $m=1$ and $m=0$ modes in Case (D-2). As in Fig.~\protect\ref{fig:E-g_D-2}(b), but with different initial conditions: (a) The perturbation amplitude is $\Psi_0 = 10^{-12}$ for all modes, and the initial phases for the five dominant modes $m=6$--$10$ are $\vartheta_{0m} = \{ 0,0,\pi,\pi,0 \}$. (b) The perturbation amplitude is $\Psi_0 = 10^{-12}$ for all modes, and the initial phases for the dominant modes $m=6$--$10$ are $\vartheta_{0m} = \{ 0,0,0,0,0 \}$. These results were obtained with a reduced number of modes, $m=0$--$31$, and are not valid far beyond $t \sim 1700$ $\tHp$. $\SHp = 10^6$, $\RHp = 10^8$.}
\label{fig:E-g_D-2_cmp-pert}%
\end{figure}

\begin{figure}
[tb]
\centering
\includegraphics[height=6.0cm,width=8.0cm]
{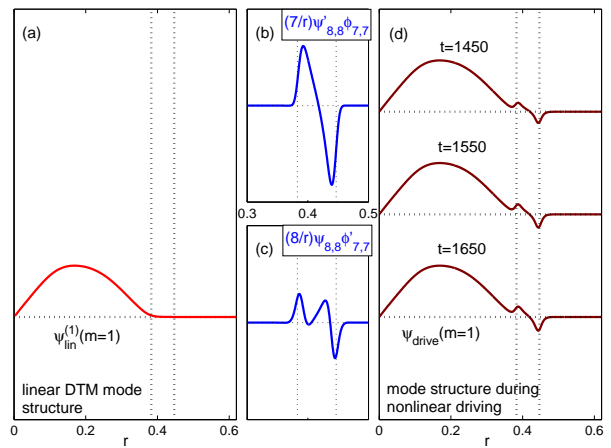}
\caption{(Color online). Effect of the nonlinear driving on the $m=1$ mode structure in Case (P-2) [cf.~Fig.~\ref{fig:E-g_D-2_cmp-pert}(b)]. The two fastest growing modes are $m=8$ and $m=7$ [cf.~Fig.~\ref{fig:spec}]. In (a) the linear eigenmode structure $\psi^{(1)}_{1,1}$ is shown. In (b) and (c) the typical structure of the driving terms can be seen. Here, the terms $(\partial_r\psi_{8,8}) (7/r)\phi_{7,7}$ and $(8/r)\psi_{8,8} (\partial_r\phi_{7,7})$ are shown, representatively for the convective nonlinearity $[\psi, \phi]$ in Eq.~(\protect\ref{eq:rmhd1}). In (d) the mode structure during the nonlinear driving phase is plotted.}
\label{fig:nl_drive}
\end{figure}

For the discussion of the evolution of individual Fourier modes we focus on Case (D-2), where two $\qRes = 1$ resonant surfaces are located a small distance apart. The results are similar in other cases with different $\qRes$, provided that $\mPeak$ is several times larger than the lowest possible mode number $m$. The evolution of the individual Fourier modes is described in terms of the kinetic and magnetic energies of their perturbation components,
\begin{equation}
\EKin_{m,n} = |\nabla\flPhi_{m,n}|^2 \quad {\rm and} \quad
\EMag_{m,n} = |\nabla\flPsi_{m,n}|^2,
\label{eq:def-E}
\end{equation}

\noindent and the corresponding nonlinear growth rates,
\begin{equation}
\gKin_{m,n}(t) = \frac{{\rm d}\ln \EKin_{m,n}}{2{\rm d}t} \quad {\rm
and} \quad \gMag_{m,n}(t) = \frac{{\rm d}\ln \EMag_{m,n}}{2{\rm d}t}.
\label{eq:def-gr}
\end{equation}

\noindent (these are amplitude growth rates, hence the factor $1/2$). In Eq.~(\ref{eq:def-E}), $|f_{m,n}|^2 \equiv \int_0^1{\rm d}r\; r\; C_m |f_{m,n}(r)|^2$, with $C_{m=0} = 4\pi$ and $C_{m\neq 0} = 2\pi$.

In Fig.~\ref{fig:E-g_D-2} the evolution of (a) the energies and (b) growth rates of the $m=1$ and $m=0$ modes in Case (D-2) is shown. Note that the $m=0$ mode considered here measures only the profile \emph{perturbation}, excluding the equilibrium profile. During phase (i) the linear mode structure of the $m=1$ mode is gradually established. In the present case, this process is not fully completed by the time the nonlinear drive (ii) sets in, as can be inferred from the fact that $\gamma_{1,1}^{\rm kin}$ and $\gamma_{1,1}^{\rm mag}$ are not equal and still vary in time.

The nonlinear driving phase (ii) begins when the fastest growing modes (not shown) reach sufficiently large amplitudes so that they start to drive slower modes through nonlinear coupling. For instance, a typical driving term in the convective nonlinearity $[\psi,\phi]$ in Eq.~(\ref{eq:rmhd1}) is
\begin{eqnarray}
&&\psi'_a(r) \sin(m_a\vartheta_*) e^{\gamma_a t} \times \phi_b(r) \sin(m_b\vartheta_*) e^{\gamma_b t}
\label{eq:coupling}
\\
&& = (\psi'_a \phi_b / 2) \left[ \cos\left(m^-\vartheta_*\right) - \cos\left(m^+\vartheta_*\right) \right] e^{(\gamma_a + \gamma_b)t},
\nonumber
\end{eqnarray}

\noindent where $m_a$ and $m_b$ are the two driving modes with linear growth rates $\gamma_a$ and $\gamma_b$, and $m^\pm = m_a\pm m_b$ are the mode numbers of driven modes. The growth rate of the driving term is $\gDrive(m^\pm) = \gamma_a + \gamma_b$.

In Fig.~\ref{fig:E-g_D-2} the $m=1$ mode switches between (almost) linear and nonlinearly driven exponential growth around $t\approx 300$. Here, the driving is primarily due to the coupling between the modes $m=8$ and $m=7$. Consequently, according to Eq.~(\ref{eq:coupling}), the expected enhanced growth rate is $\gDrive(m=1) \approx \gDrive(m=0) = 2\gLin(m=8) = 17.5\times 10^{-3}$. The level of nonlinear driving can conveniently be inferred from the growth rate of the $m=0$ mode (only magnetic energy). The $m=0$ mode is not an unstable eigenmode and its evolution is entirely a result of nonlinear driving by higher-$m$ modes.

The annular collapse, labeled by (iii) in Fig.~\ref{fig:E-g_D-2}, begins already during the driving phase and continues beyond it. The dynamics during this stage were described above in Section~\ref{sec:results_islands}.

In Fig.~\ref{fig:E-g_D-2} neither the $m=0$ nor the $m=1$ mode reach the expected growth rate, $\gDrive = 17.5\times 10^{-3}$. The reason for this lies in the initial perturbation. We have determined two factors that need to be considered: (A) the perturbation amplitude, and (B) the phase relations between the driving modes.

The effect of (A) can be seen by comparing Fig.~\ref{fig:E-g_D-2}(b) with Fig.~\ref{fig:E-g_D-2_cmp-pert}(a). Figure~\ref{fig:E-g_D-2_cmp-pert} shows results for Case (D-2) that were obtained with a lower perturbation amplitude $\Psi_0=10^{-12}$ [in Fig.~\ref{fig:E-g_D-2}: $\Psi_0 = 10^{-7}$]. Now, there is more time for all modes to establish the linear mode structures. Clearly, the nonlinear driving in Fig.~\ref{fig:E-g_D-2_cmp-pert}(a) reaches a higher level than in Fig.~\ref{fig:E-g_D-2}(b).

Factor (B) implies that the growth rate $\gDrive$ may also stay below $2\gPeak$ when there are several higher-$m$ modes with growth rates approximately equal to $\gPeak$ and when the phase relations between these modes are ``unfavorable.'' Thus, $\gDrive$ can be increased by ``aligning'' the phases of the fastest growing modes in the spectrum. The result of aligning the five modes $m=6$--$10$ is shown in Fig.~\ref{fig:E-g_D-2_cmp-pert}(b). Clearly, the driving of the $m=1$ mode is now much more effective.

Strictly speaking, the effect of aligning the relative phases is merely to reduce the time needed to establish the nonlinear driving. Its effect on the growth rate is only temporary. Note that the growth of the $m=0$ mode is not affected by phase relations between the driving modes, as is to be expected.

The effect of the nonlinear driving on the $m=1$ mode structures can be observed in Fig.~\ref{fig:nl_drive}. Note in particular that despite the radial localization of the driving terms [Fig.~\ref{fig:nl_drive}(b) and (c)], the $m=1$ mode [Fig.~\ref{fig:nl_drive}(a)] as a \emph{whole} grows at an enhanced rate. Once the driving is established the mode structure varies only minutely [Fig.~\ref{fig:nl_drive}(d)]. Thus, a new nonlinear mode structure is formed. One consequence of this global effect of the localized driving is that in cases where the driving term happens to have the opposite sign, it induces a switching of the sign of the global $m=1$ mode structure at the onset of the nonlinear driving phase. Such an event can be observed in Fig.~\ref{fig:E-g_D-2_cmp-pert}(b), where $\EMag_{2,1}$ performs an under- and overshoot as a result of sign reversal.

\section{Discussion and Conclusions}
\label{sec:conclude}

In this paper we have studied
the linear instability and nonlinear dynamics of resistive DTMs for small inter-resonance distance $D_{12}$. The results may be summarized as follows.

The linear growth rates $\gLin(m)$ of low-$m$ modes were found to increase with $m$. The power law $\gLin \propto m^{2/3}$ predicted by the linear theory developed in Ref.~\cite{Pritchett80} agrees with numerical results. The linear growth rates of high-$m$ modes decrease with $m$. These modes seem to be outside the scope of the existing theory. A semi-empirical formula for the mode number of the fastest growing mode $\mPeak$ was proposed [Eq.~(\ref{eq:mpeak})], based on results of Ref.~\cite{Pritchett80}. The estimate for $\mPeak$ is valid only for \emph{resistive} DTMs. Nevertheless, broad spectra and high values $\mPeak$ are also found when reconnection is mediated by electron inertia instead of resistivity, as is shown in a separate paper \cite{Bierwage06d}. Also, Eq.~(\ref{eq:mpeak}) does not apply to cases with more than two resonant surfaces. For instance, for triple tearing modes (TTMs) both $\mPeak$ and the number of unstable modes are larger than for DTMs under comparable conditions, indicating that a TTM tends to be an even stronger instability \cite{BierwageThesis}.

The dominance of several high-$m$ DTMs in configurations with small $D_{12}$ leads to an annular collapse with small island structures. This turbulent annular collapse shows that the nonlinear DTM dynamics in cases with small inter-resonance distances (dominant high-$m$ modes) are different from those studied previously by other authors using large inter-resonance distances (dominant lowest-$m$ mode). Our results may be of interest for scenarios, where $\qMin$ is gradually lowered by increasing the off-axis current. Experimentally, an apparently quiescent passage through low-order resonant surfaces and thus access to regimes with larger inter-resonance distance, was found to be possible by applying additional external drive (e.g., Ref.~\cite{Guenter99}). MHD activity was reported after the inter-resonance distance had grown substantially and the dynamics seem to be dominated by low-$m$ modes. It is possible that earlier high-$m$ activity predicted by our simulations is either prevented by the external drive (not included in our model) or has escaped detection. Through recent progress in plasma diagnostics, which allows to detect high-$m$ magnetic islands on low-order resonant surfaces \cite{Donne05}, experimental checks of our simulation results may be feasible in the near future.



Before entering the fully nonlinear regime, the fastest growing modes drive slower modes. For $\qRes = 1$, this includes the \emph{global} $m=1$ resistive internal kink mode, despite the fact that the driving is radially \emph{localized} between the resonant surfaces. This nonlinear driving experienced by slower modes is of particular interest for the $m=1$ internal kink mode. The nonlinear driving implies that the $m=1$ mode appears significantly earlier than would be expected from its linear growth rate. Hence, the fast growing high-$m$ DTMs effectively provide a trigger mechanism for the $m=1$ mode, similarly to the TTM case described in detail in Ref.~\cite{Bierwage06a}. Note that, by the time the $m=1$ mode becomes observable, the inter-resonance region has already undergone reconnection. The lack of magnetic shear allows the $m=1$ mode to grow as an \emph{ideal} internal kink instability. In contrast to the TTM case, there is no sawtooth crash associated with the activity of the $m=1$ mode when the inter-resonance distance is small. However, the rapid excitation of $m=1$ oscillations observed in our simulations may be useful in experiments to determine the instant in time when $\qMin$ passes through the $\qRes = 1$ resonant surface.

It is important to note that the instability of a broad spectrum of modes implies that the \emph{details} of the nonlinear dynamics depend on the initial conditions used in the simulation. For instance, the efficiency of the nonlinear driving was shown to depend on the phase relations between the fastest growing modes (cf.~Fig.~\ref{fig:E-g_D-2_cmp-pert}). Furthermore, the phase relations determine where the first magnetic islands form. In principle, it is possible to produce the first magnetic islands in a poloidally localized region. Although, this discussion of initial perturbations may seem rather academic, we believe that it may bear practical importance. For instance, pellet or neutral beam injection may provide a localized magnetic perturbation and possibly form robust helical structures such as ``snakes'' \cite{Wesson95}. Furthermore, if the DTMs are excited by micro-turbulence, modes with $m > \mPeak$ may dominate the annular collapse, in contrast to the present study, where all modes were perturbed with similar amplitude so that $\mPeak$ became dominant.

In conclusion, the results presented in this paper may be relevant to the understanding of the MHD activity near $\qMin$ and may thus be of interest for studies on stability and confinement of advanced tokamaks. The results motivate further research in this direction with more realistic models.

\begin{acknowledgments}
A.B. would like to thank Y. Kishimoto, Y. Nakamura and F. Sano for valuable discussions. S.B. acknowledges the Graduate School of Energy Science at Kyoto University and the Center for Atomic and Molecular Technologies at Osaka University for their support and hospitality under visiting professorships. S.H. acknowledges the University of Provence for support and hospitality under visiting professorship. This work was partially supported by the 21st Century COE Program at Kyoto University and the Japan-France Integrated Action Program ``SAKURA'' for Osaka University and the University of Provence.
\end{acknowledgments}


\end{document}